# Stimulated low-frequency Raman scattering in tobacco mosaic virus suspension

O V Karpova[1], A D Kudryavtseva[2], V N Lednev[3], T V Mironova[2], V B Oshurko[4], S M Pershin[3], E K Petrova[1], N V Tcherniega[2] and K I Zemskov[2]

[1]M.V. Lomonosov Moscow State University, Vorob'evy Gory, 1, Moscow, 119991, Russia
[2]P.N. Lebedev Physical Institute of the RAS, Leninskii pr, 53, Moscow, 119991, Russia
[3]A.M. Prokhorov General Physics Institute of the RAS, Vavilova, 38, Moscow, 119991, Russia
[4]Moscow Technical University "STANKIN", Vadkovskii per. 3ª, Moscow, 127055, Russia

**Abstract**. Laser pulses interaction with tobacco mosaic virus (TMV) in Tris-HCl pH7.5 buffer and in water has been investigated. 20 ns ruby laser pulses have been used for excitation. Spectrum of the light passing through the sample was registered with the help of Fabri-Perot interferometer. In the case of TMV in water we observed in the spectrum only one line of the exciting laser light, for TMV in Tris-HCl pH7.5 buffer second line appeared, corresponding to the stimulated low-frequency Raman scattering (SLFRS) on the breathing radial mode of TMV. SLFRS frequency shift by 2 cm$^{-1}$, (60 GHz), conversion efficiency and threshold are measured for the first time to the best of our knowledge.

**Keywords**: *nanoparticle, stimulated scattering, vibration, laser, virus*.

PACS: 42.65.Dr

**Introduction**

Low frequency Raman scattering (LFRS) [1,2] is a powerful tool for submicron and nanoparticles systems investigations. The LFRS spectrum analysis gives unique possibility to get the information about morphological properties of the system under consideration including determination of the size distribution [3].

Currently LFRS is realized for systems containing metallic, insulator, or semiconductor nanoparticles [3]. Biological nanoparticles (including viruses) are also the subject of intensive theoretical and experimental research [4-6]. Viruses of cylindrical or spheroidal shape with their own acoustic vibrations eigenfrequencies are a good example of highly monodisperse nanoparticles system. LFRS in system of cylindrical shape viruses was observed in [7].

Analysis of the low frequency spectrum of the inelastically spontaneous scattered light in rigid biological structures like in any nanoparticles system can give very important information about their mechanical properties and can be used for their identification. But compared with metal, dielectric or semiconductor nanoparticles acoustic properties of most biological nanoparticles are not well defined making it difficult to analyze the experimental results. Another very important problem with LFRS investigation in different viruses is damping of the vibration due to acoustic energy radiation into the environment particularly for the case of water. This process can lead to strong intensity reduction for some LFRS modes [6,7]. Influence of the liquid environment viscosity both on frequency and vibrational modes damping was considered in [8].

Like any spontaneous scattering LFRS has its stimulated analogue – stimulated low-frequency Raman scattering (SLFRS) [9,10]. SLFRS main features unlike its spontaneous analogue are high conversion efficiency of the exciting light into the first Stokes component, small divergence of the scattered light propagating in forward and backward directions, narrow spectral distribution as well as the presence of threshold. The relation between LFRS and SLFRS is nearly the same as between spontaneous and stimulated Raman scattering. In this Letter we present the results of the first to the best of our knowledge experimental investigation of SLFRS excited in the suspension of tobacco mosaic virus (TMV) in Tris-HCl pH7.5 buffer and in water also.

**Experiment**

The rod-like particles of tobacco mosaic virus (TMV) of 18 nm diameter and 300 nm modal length consist of 2130 identical 17.5 kDa protein subunits arranged helically into a rigid tube. The viral RNA is intercalated between the protein turns [11]. TMV strain U1 was isolated from systemically infected Nicotiana tabacum L. cv. Samsun plants as described previously [12]. The TMV concentration in the test sample was 50 μg/ml in Tris-HCl pH7.5 buffer. The number of particles in the sample was analyzed by Nanoparticle tracking analysis (NTA) according to [13,14]. The TMV concentration in the sample used was $0.5 \times 10^{12}$ particles/cm$^{-3}$. Also the sample of TMV in water with the same concentration was used.

For SLFRS excitation ruby laser ($\lambda$ = 694.3 nm, $\tau$ = 20 ns, $E_{max}$ = 0.3 J, $\Delta\nu$ = 0.015 cm$^{-1}$, divergence 3.5·10$^{-4}$ rad) was used. Laser light was focused at the center of the 1 cm quartz cell with sample by the lens with focal length 5 cm. SLFRS spectra have been registered with Fabri-Perot interferometers with the range of dispersion 2.5 cm$^{-1}$ (75 GHz). For comparison the same quartz cell filled only with Tris-HCl pH7.5 buffer was used as a reference sample.

Experimental setup for SLFRS investigations in TMV suspensions is shown in the Figure 1.

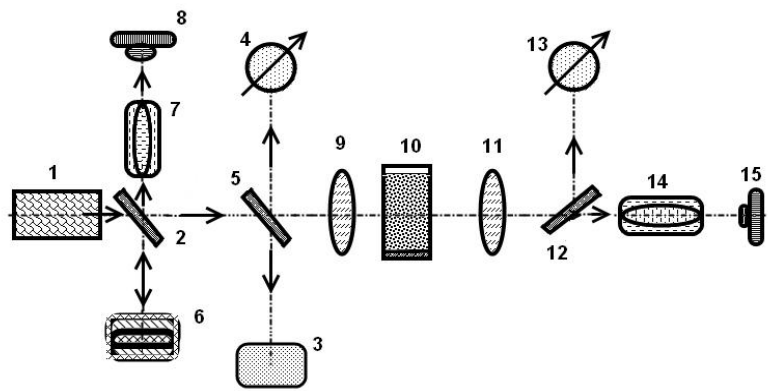

**Figure 1.** Experimental setup: 1 – ruby laser; 2, 5, 12 – glass plates; 3 – system for laser pulse characteristics measurements; 4, 13 – systems for SLFRS energy measuring in forward and backward direction; 6 – mirror; 7, 14 – Fabri-Perot interferometers; 8, 15 - photo cameras, registering SLFRS spectra; 9, 11 – lenses, 10 – quartz cell with the sample.

Simultaneously with spectral measurements the radiation energy transmitted through and reflected from the cell was measured by calibrated photodiodes. For relative SLFRS intensity definition the blackening marks were used.

**Results and discussion**

Spectrum of light scattered from the cell filled with TMV suspension in Tris-HCl pH7.5 buffer in forward and backward directions was measured simultaneously. At the laser pulse low level intensities only one system of rings was registered which corresponds to the laser light frequency (Figure 2a). Increasing laser intensity leads to the SLFRS excitation. At the laser pulse intensity exceeding a certain threshold SLFRS both in forward and backward directions was registered (Figure 2b).

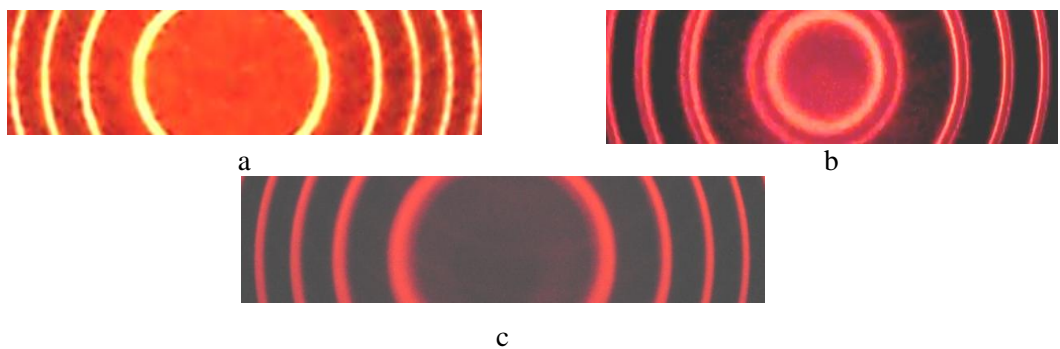

**Figure 2.** Fabri-Perot interferograms corresponding to the forward scattered radiation in TMV suspension in Tris-HCl pH7.5 buffer for laser intensity a) 0.02 GW/cm$^2$, b) 0.08 GW/cm$^2$ ; c) interferogram in the Tris-HCl pH7.5 buffer for laser intensity 0.08 GW/cm$^2$.

Additional system of rings on the interferogram corresponds to the scattered light. In the same range of laser intensities no scattering in the cell filled only with Tris-HCl pH7.5 buffer was registered.

For the experimental conditions the threshold value of SLFRS excitation was 0.07 GW/cm$^2$. The threshold for backward and forward scattered light was approximately the same. Line width and divergence of SLFRS were nearly the same as the corresponding values of the laser light. Maximum conversion efficiency for SLFRS scattered in forward direction was about 5 per cent at the experimental conditions of excitation.

SLFRS frequency shift was found to be 2 cm$^{-1}$ (60 GHz). The same frequency shift was registered both for forward and backward scattered waves. Simultaneous excitation of the scattering in opposite directions is an exact evidence of acoustic excitation of the localized modes corresponding to the TMV eigenfrequencies. According to [6] the LFRS frequency shift value for radial breathing mode is 1.85 cm$^{-1}$ for TMV in air and 2.1 cm$^{-1}$ for TMV virus in water. Our experimentally measured value 2 cm$^{-1}$ is quite close to the calculated one [6]. It is necessary to point out that the SLFRS frequency shift is defined by the size and shape of the virus and environment properties so the value of this frequency shift can be used for identification of the system under consideration. The presence of the scattering threshold, high conversion efficiency, sharp and narrow spectral line, small beam divergence – these are the main features of SLFRS excited in TMV Tris-HCl pH7.5 buffer suspension.

At our experimental conditions SLFRS in TMV in water was not excited. It can be connected with modes damping due to radiation of the acoustic energy into the environment [8]. This leads to an increase in the threshold for SLFRS excitation which exceeds the laser intensity we used. In contrast, using Tris-HCl pH7.5 buffer instead of water leads to acoustic impedance mismatch at the virus surface (less damping) and thus lowers the SLFRS threshold for excitation by 20 ns pulse duration.

Comparison of the transmission electron microscopy images of TMV before and after measurements showed that TMV morphology was not changed after multiple laser irradiations of the sample (Figure 3). Sample shown in Fig. 3B was irradiated by 20 laser pulses with intensity above the SLFRS threshold. SLFRS conversion efficiency was about maximum value (5%) for all 20 laser pulses.

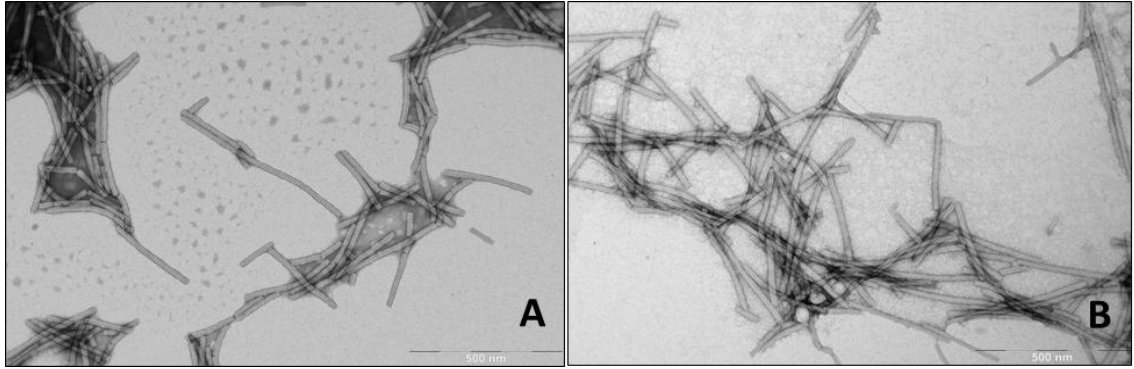

**Figure 3.** Transmission electron microscopy image of the samples of the virus before (**A**) and after sample irradiation by 20 laser pulses (**B**).

**Conclusions**

The generation of SLFRS from virus suspension is accompanied by intense virus vibrations. Due to high conversion efficiency and monochromaticity of the scattered light virus vibrations are spatially and temporally coherent. Coherent buildup of viruses' vibrations mechanism is analogous to those which take place for coherently driven lattice or molecular vibrations in the process of the stimulated Raman scattering [15] and can be briefly described in the next way. The powerful electromagnetic laser field induces the polarization of the viruses which are vibrating with their eigenfrequency. This polarization is the source of the inelastically scattered wave. Interaction of both waves produces a force through induced polarization. This mechanical force coherently excites Raman-active vibrational modes of the virus. High SLFRS conversion efficiency is exact evidence of effective coherent excitation of viruses' vibrations like in suspensions of silver and gold nanoparticles. If necessary the optical set up allows realizing the effective SLFRS amplification using feedback with the laser resonator. So SLFRS in TMV and other virus suspension at our experimental conditions can be used as the source of biharmonic pumping for investigating the systems with eigenfrequencies in gigahertz range and also for powerful impact on such systems. This item will be the subject of the following paper.

**Acknowledgments**

This work was partly supported by Russian Foundation for Basic Research Grant 14-02-00748-a and Russian Ministry of Science and Education Grant # 1678.

**References**

1. Duval E, Boukenter A and Champagnon B 1986 Vibration Eigenmodes and Size of Microcrystallites in Glass: Observation by Very-Low-Frequency Raman Scattering *Phys. Rev. Lett*. **56** 2052-2055.
2. Ivanda M, Babocsi K, Dem C, Schmitt M, Montagna M and Kiefer W 2003 Low-wavenumber Raman scattering from CdSxSe1-x quantum dots embedded in a glass matrix *Phys. Rev. B* **67** 235329.
3. Montagna M 2008 Brillouin and Raman scattering from the acoustic vibrations of spherical particles with a size comparable to the wavelength of the light *Phys. Rev. B* **77** 045418.
4. Talati M, Jha P K 2006 Acoustic phonon quantization and low-frequency Raman spectra of spherical viruses *Phys. Rev. E*, **73**, 011901.
5. Dykeman E C, Sankey O F and Kong-Thon Tsen 2007 Raman intensity and spectra predictions for cylindrical viruses *Physical Review E* **76**, 011906.
6. Balandin A and Fonoberov V 2005 Vibrational Modes of Nano-Template Viruses *Journal of Biomedical Nanotechnology* **1** 90-95.
7. Tsen K T, Dykeman E C, Sankey O F, Lin N T, Tsen S W D and Kiang J G 2006 *Virology Journal* **3**, 79.
8. Murray D B and Saviot L 2007 Damping by bulk and shear viscosity for confined acoustic phonons of a spherical virus in water *Journal of Physics: Conference Series* **92**, 012036.
9. Tcherniega N V, Samoylovich M I, Kudryavtseva A D, Belyanin A F, Pashchenko P V and Dzbanovski N N 2010 Stimulated scattering caused by the interaction of light with morphology-dependent acoustic resonance *Optics Letters* **35** 300-302.
10. Tcherniega N V, Zemskov K I, Savranskii V V, Kudryavtseva A D, Olenin A Yu and Lisichkin G V 2013 Experimental observation of stimulated low-frequency,Raman scattering in water suspensions of silver and gold nanoparticles *Optics Letters* **38** 824-826.
11. Klug A 1999 The tobacco mosaic virus particle: structure and assembly *Philos Trans R Soc Lond B Biol Sci.* **354** 531–535.
12. Karpova O, Nikitin N, Chirkov S, Trifonova E, Sheveleva A, Lazareva E and Atabekov J 2012 Immunogenic compositions assembled from tobacco mosaic virus-generated spherical particle platforms and foregn antigens *Journal of General Virology* **93** 400–407.

13. Nikitin N, Trifonova E, Karpova O and Atabekov J. 2013 Examination of Biologically Active Nanocomplexes by Nanoparticle Tracking Analysis *Microscopy and Microanalysis* **19** 808-813.

14. Petrova E, Nikitin N, Trifonova E, Protopopova A, Karpova O and Atabekov J 2015 The 5'-proximal region of Potato virus X RNA involves the potential cap-dependent "conformational element" for encapsidation *Biochimie* **115** 116-119.

15. Garmire E, Pandarese F and Towns C H 1963 Coherently driven molecular vibrations and light modulation *Phys. Rev. Letters* **11** 110-163.